\documentclass[conference]{IEEEtran}

\makeatletter
\def\ps@headings{%
\def\@oddhead{\mbox{}\scriptsize\rightmark \hfil \thepage}%
\def\@evenhead{\scriptsize\thepage \hfil \leftmark\mbox{}}%
\def\@oddfoot{}%
\def\@evenfoot{}}
\makeatother
\usepackage{cite}
\usepackage{amsmath,amssymb,amsfonts}
\usepackage{graphicx}
\usepackage{textcomp}
\usepackage{xcolor}
\usepackage{adjustbox}

\usepackage{multirow}
\usepackage{stfloats}
\usepackage{booktabs}
\usepackage{color, colortbl}
\usepackage{subcaption}
\usepackage{algorithm}
\usepackage{algpseudocode}
\usepackage{diagbox}
\usepackage{seqsplit}
\usepackage{soul}

\definecolor{red}{HTML}{de2d26}
\definecolor{green}{HTML}{31a354}

\AtBeginDocument{%
  \providecommand\BibTeX{{%
    Bib\TeX}}}

\def\BibTeX{{\rm B\kern-.05em{\sc i\kern-.025em b}\kern-.08em
    T\kern-.1667em\lower.7ex\hbox{E}\kern-.125emX}}

\begin{document}

\title{Benchmarking Android Malware Detection: Traditional vs. Deep Learning Models}

\author{\IEEEauthorblockN{Guojun Liu}
  \IEEEauthorblockA{
    University of South Florida\\
    Tampa, Florida\\
    guojunl@usf.edu}
\and
\IEEEauthorblockN{Doina Caragea}
\IEEEauthorblockA{
  Kansas State University\\
  Manhattan, Kansas\\
  dcaragea@ksu.edu}
\and
\IEEEauthorblockN{Xinming Ou}
  \IEEEauthorblockA{
    University of South Florida\\
    Tampa, Florida\\
    xou@usf.edu}
\and
\IEEEauthorblockN{Sankardas Roy}
\IEEEauthorblockA{
  Bowling Green State University\\
  Bowling Green, Ohio\\
  sanroy@bgsu.edu}
}

\maketitle

\begin{abstract}

  Android malware detection has been extensively studied using both
traditional machine learning (ML) and deep learning (DL)
approaches. While many state-of-the-art detection models, particularly
those based on DL, claim superior performance, they often rely on
limited comparisons, lacking comprehensive benchmarking against
traditional ML models across diverse datasets. This raises concerns
about the robustness of DL-based approaches' performance and the
potential oversight of simpler, more efficient ML models. In this
paper, we conduct a systematic evaluation of Android malware detection
models across four datasets: three recently published, publicly
available datasets and a large-scale dataset we systematically
collected. We implement a range of traditional ML models, including
Random Forests (RF) and CatBoost, alongside advanced DL models such as
Capsule Graph Neural Networks (CapsGNN), BERT-based models, and
ExcelFormer-based models. Our results reveal that 
in many cases simpler and more computationally efficient ML models achieve
comparable or even superior performance compared with DL models. These findings highlight the
need for rigorous benchmarking in Android malware detection
research. We encourage future studies to conduct more comprehensive
benchmarking comparisons between traditional and advanced models to
ensure a more accurate assessment of detection capabilities. To
facilitate further research, we provide access to our dataset,
including app IDs, hash values, and labels.

\end{abstract}


\begin{IEEEkeywords}
Android malware detection, Machine learning, Deep learning, Dataset, Benchmark
\end{IEEEkeywords}

\section{Introduction}
The Android ecosystem maintains the leading position as the largest
and most active system among the worldwide mobile computing
environment by far. 
According to one recent {\it 42matters}'s
statistics~\cite{42matters:googleplay2025}, there were 40,325 apps on
average released monthly through Google Play alone between
November 2024 and Juanuary 2025.  The app market’s rapid growth makes
it crucial to have an automated and effective app vetting system that
scales to the app market sizes.  An important component of a vetting
system is the application of machine learning to triage the apps in
the vetting process.
Researchers train a machine learning model on a dataset consistsing of
apps whose maliciousness is known. Then the trained model is applied to
a disjoint test dataset to evaluate the model's effectiveness.

Traditional machine learning approaches (shorthanded ML hereafter)
such as Naive Bayes (NB), k-Nearest Neighbors (k-NN), Support Vector
Machines (SVM), Random Forest (RF), Artificial Neural Networks (ANN),
Gradient Boosting, and so on,  have
shown outstanding performance~\cite{Iker:SPSM11, Aafer:SecureComm13,
Gascon:AISec13, WiSec:Saurabh13, Arp:NDSS14, Yang:ESORICS14,
Periravian:ICTAI13, Chen:ASIACCS16, Yan:MobiSys19, Gong:EuroSys20} on
datasets from small to market-scale.
More recently, deep learning approaches (shorthanded DL hereafter)
have been studied on Android malware detection~\cite{Yuan:SIGCOMM14,
Su:TrustCom16, Hou:WIW16, McLaughlin:CODASPY17, Karbab:DFRWS18,
Xu:EuroS&P18, Xu:ICFEM18, Xiao:MTA19, Kim:TIFS18, Wang:JAIHC19,
Oak:AISec19, Alzaylaee:Computer&security20, Nadia:MALHat21,
Chen:USENIX23, Sun:esem24}. One potential advantage of DL over ML is
that it does not require an extensive feature selection and
engineering process, although careful consideration still needs to be
applied as to how to represent app data provided as input to the DL
models.

Despite various DL approaches proposed for Android malware detection,
there is limited comparison between DL and traditional ML
models~\cite{Yuan:SIGCOMM14, Wang:JAIHC19,Shar:MOBILESoft20,
Nadia:MALHat21, Sun:esem24}. Evaluating DL models with only a few
baselines often fails to reflect true performance -- a pitfall
termed ``inappropriate baseline''~\cite{Daniel:USENIXSecurity22}
previously noted in areas like vulnerability discovery and network
intrusion detection. We observed similar shortcomings in Android
malware detection research. Many studies do not release their
datasets, making it difficult to benchmark against a broader range of
alternatives. Encouragingly, some recent works have made their
datasets publicly available. Through our literature review, we
identified two studies that 1) compared DL and ML models, 2) published
their datasets, and 3) used datasets of substantial sizes. This allowed
us to further benchmark those DL models against additional ML
models. We also evaluated two other DL models against ML models using
a large dataset we curated. Our experiments show mixed results
regarding whether DL consistently outperforms ML models, underscoring
the need for more comprehensive benchmarking in Android malware
detection.

\section{Related Work}
\label{sec:related}

Machine learning-based Android malware detection has been studied for over a decade. Early approaches like Crowdroid~\cite{Iker:SPSM11} used k-means clustering to separate benign apps from malware, DroidAPIMiner~\cite{Aafer:SecureComm13} extracted static API-level features and applied classifiers such as k-NN, SVM, and Decision Trees (DT). DREBIN~\cite{Arp:NDSS14} represented apps using a rich static feature set and classified them using SVM. MaMaDroid~\cite{Mariconti:NDSS2017} abstracted API calls into Markov chains and used classifiers like RF and SVM. APIChecker~\cite{Gong:EuroSys20} collected dynamic API features and evaluated nine ML algorithms, with RF performing best on a dataset of approximately 500K apps. Other systems, including DroidMat~\cite{Wu:AsiaJCIS12}, APIGraph~\cite{Zhang:CCS20}, and MalScan~\cite{Wu:ASE19}, employed lightweight static or graph-based features with classifiers like k-NN and RF, consistently demonstrating strong detection performance. 

Recent research has increasingly focused on deep learning (DL) for Android malware detection, leveraging models such as deep belief networks (DBN)~\cite{Yuan:SIGCOMM14, Su:TrustCom16}, convolutional neural networks (CNN)~\cite{McLaughlin:CODASPY17, Karbab:DFRWS18,Nadia:MALHat21, Arslan:CCPE22}, long short-term memory (LSTM) networks~\cite{Vinayakumar:JIFS18, Xiao:MTA19}, pre-trained BERT models~\cite{Sun:esem24}, and multilayer perceptrons (MLP)~\cite{Chen:USENIX23}. Droid-Sec~\cite{Yuan:SIGCOMM14} used DBN and showed improved accuracy over traditional ML models. Deep4MalDroid~\cite{Hou:WIW16} applied stacked autoencoders (SAEs) and outperformed classifiers like SVM and NB. DeepFlow~\cite{Zhu:ISCC17} showed DBN and SVM achieved better balance between recall and precision than other ML models. DexRay~\cite{Nadia:MALHat21} transformed DEX bytecode into grayscale images for CNN-based detection, achieving performance comparable to DREBIN. DetectBERT~\cite{Sun:esem24} used BERT~\cite{bert:naacl19} with Multiple Instance Learning to process Smali code, improving class-level representation. DeepRefiner~\cite{Xu:EuroS&P18} combined MLP and LSTM in a two-layer system, achieving ~10\% higher accuracy than SVM. DANdroid~\cite{Millar:CODASPY20} evaluated a discriminative adversarial network on the DREBIN dataset, showing similar results. DL-Droid~\cite{Alzaylaee:Computer&security20} extracted 420 dynamic features and used MLP, outperforming seven ML baselines. AMD-CNN~\cite{Arslan:CCPE22} visualized Android manifest features as RGC images for CNN input. Chen et al.~\cite{Chen:USENIX23} proposed a continuous learning MLP model for handling concept drift~\cite{Jordaney:USENIX17}, evaluated on APIGraph~\cite{Zhang:CCS20} and AndroZoo~\cite{Allix:MSR16} with major baseline models.

\section{Benchmarking Methodology}
\label{sec:benchmark_method}

Benchmarking across different models is a standard approach for
evaluating and comparing their performance, as well as identifying the
strengths and weaknesses of various approaches. However, determining
the sufficient number of models to compare with is not easy, as
further benchmarking using more models could always be beneficial. For
this reason, it is utterly important that researchers make the dataset
upon which their models are evaluated available. This will enable
other researchers to evaluate additional models that had not been
considered in the original research. More extensive evaluations can
provide deeper insights and help understand the true strengths of the
various models for Android malware detection.

Our primary objective is not to assert the superiority of any specific
model but to demonstrate how comprehensive benchmarking enhances the
understanding of model performance. To this end we adopt the following
benchmarking strategy.

\begin{enumerate}
\item We keep the app representation (e.g., feature vectors) used in
the original research.  When the provided datasets include feature
vectors used in learning, we simply use that information. When the
original feature vectors are not publicly available, we generate them
from the App apk files based on the information provided in the paper.

\item We use the exact same datasets as used in the original research,
and benchmark additional ML models that were not examined in the
original research.
\end{enumerate}

Since each work uses different datasets, the second point indicates
that when interpreting the results presented in this paper,
comparisons shall only be made within each work. Comparison of
performance results across different works is not meaningful.  While
it may be tempting to use the same dataset to evaluate all the models
proposed in all these works, the research community has not agreed
upon a ``gold standard'' dataset for Android malware detection.  In
fact to understand how choices in datasets construction impact models'
measured performance is a non-trivial research question in its own
right and beyond the scope of this paper.

\subsection{Traditional Machine Learning Models Considered}
\label{subsec:traditional_ml}

Widely explored ML models in Android malware detection include Naive
Bayes (NB)~\cite{Yerima:AINA13, Sharma:CANS14}, support vector
machines (SVM)~\cite{Gascon:AISec13, Arp:NDSS14}, k-nearest neighbors
(k-NN)~\cite{Wu:AsiaJCIS12, Sharma:CANS14, Wu:ASE19}, and random
forest (RF)~\cite{Mariconti:NDSS2017, Wu:ASE19}.  We consider these
four widely used models as potential baselines to compare with the
proposed DL models.  Additionally, Gradient Boosted Decision Trees
(GBDTs) are powerful classification tools, with
CatBoost~\cite{CatBoost:NEURIPS2018} being a notable variant of
GBDT-based ensemble techniques.  Having been introduced in late 2018,
CatBoost has demonstrated strong performance in classification and
regression tasks across various domains, including
cybersecurity~\cite{CatBoost:NEURIPS2018}. Thus we also include it as
one of the ML models in our benchmarking.

\subsection{Deep Learning Models Considered}
\label{subsec:dl_models}

One of the key challenges in applying recent DL methods to Android
malware detection is the lack of publicly available datasets in their
studies.  Fortunately, we have identified recent studies, such as Chen
et al.’s Enc+MLP~\cite{Chen:USENIX23} and Sun et al.’s
DetectBERT~\cite{Sun:esem24}, that have made their datasets publicly
available. This allowed us to conduct more extensive benchmarking of the
models proposed in these two works.  Additionally, we collected a
Google Play-only dataset and used two DL approaches, Capsule Graph
Neural Networks (CapsGNN) and ExcelFormer, which have not been
explored before for Android malware detection.  Each of the three
works employs different datasets to evaluate their models, and their
approaches to app representation also vary.

\section{Case Studies}
\label{sec:case_studies}

\subsection{Case Study 1: \textbf{DetectBERT}}
\label{subsec:case_detectbert}

DetectBERT is a recent approach proposed by Sun et al.~\cite{Sun:esem24}, which utilizes a pre-trained BERT-like model combined with Correlated Multiple Instance Learning (c-MIL) to process Smali code from APKs. This approach enhances class-level features and aggregates them into an app-level representation, improving the effectiveness of Android malware detection. 

\subsubsection{Benchmarking for DetectBERT}
DetectBERT was evaluated against two state-of-the-art Android malware
detection models, Drebin~\cite{Arp:NDSS14} and
DexRAY~\cite{Nadia:MALHat21}. Drebin utilizes SVM on static feature
vectors, while DexRAY is a DL framework that transforms low-level
bytecode into images and processes them using a CNN for malware
detection. The results indicate that DetectBERT slightly outperforms
both baseline models.

\subsubsection{Dataset for DetectBERT}
\textbf{DetectBERT Benchmark Dataset.} This large-scale dataset originates from the DexRay~\cite{Nadia:MALHat21} study and comprises 96,994 benign apps and 61,809 malware apps. The labeling process is based on VirusTotal reports: apps that are not flagged by any antivirus engines are considered benign, while those detected as malicious by more than two antivirus engines are labeled as malware.  

\subsubsection{Additional Benchmarking}

Sun et al.~\cite{Sun:esem24} open sourced their DetectBERT model for
Android malware detection, which enables us to use the exact same
dataset, data splits, and evaluation metrics for training and testing
various models, ensuring fair comparisons.

One of the challenges in applying the DetectBERT model is generating
the feature vector for each smali class in an APK file from a
large-scale dataset. To address this, we explored the performance of
more cost-effective approaches on this dataset.  Since DREBIN is one
of the baseline models used in the original research, we generated
DREBIN-like static features for each app, including requested
permissions, component names, intents, suspicious API calls, and
network addresses. We then calculated the mutual information of the
unique static features in this dataset and ranked them accordingly. By
testing different feature vector sizes based on the top N features, we
found that a feature vector consisting of 2,919 static features per
app yielded optimal results.  Table~\ref{tab:detectbert_comparison}
presents the performance of various ML models against DetectBERT. The
shaded rows represent the models we tested with these features. Except
for NB, all other models (KNN, SVM, RF, and CatBoost) achieved
performance comparable to DetectBERT. CatBoost slightly outperforms
DetectBERT.  These ML models also require significantly less compute
time than DetectBERT.

\begin{table}[t]
  \centering
  \resizebox{0.45\textwidth}{!}{
    \renewcommand{\arraystretch}{0.9}
    \begin{tabular}{lcccc}
        \toprule[0.8pt]
        \textbf{Model} & \textbf{Accuracy} & \textbf{Precision} & \textbf{Recall} & \textbf{F1 Score} \\
        \hline
        Drebin & 0.97 & 0.97 & 0.94 & 0.96 \\
        DexRay & 0.97 & 0.97 & 0.95 & 0.96 \\
        \textbf{DetectBERT} & \textbf{0.97} & \textbf{0.98} & \textbf{0.95} & \textbf{0.97} \\

        \midrule
        \rowcolor{gray!20} NB & 0.85 & 0.87 & 0.85 & 0.86 \\
        \rowcolor{gray!20} RF & 0.97 & 0.97 & 0.97 & 0.97 \\
        \rowcolor{gray!20} KNN & 0.97 & 0.97 & 0.97 & 0.97 \\
        \rowcolor{gray!20} SVM & 0.96 & 0.96 & 0.96 & 0.96 \\
        \rowcolor{gray!20} CatBoost & 0.98 & 0.98 & 0.98 & 0.98 \\
        \bottomrule[0.8pt]
    \end{tabular}
    }
    \caption{Model performance on the DetectBERT benchmark dataset. Unshaded results are from the original paper; shaded results are from our benchmarking.}
    \label{tab:detectbert_comparison}
  \end{table}

In the paper the authors also conducted experiments on data sets that
adhere to temporal consistency. However the published dataset does not
include the train/test data split for this experiment, and as a result
we were not able to compare additional ML models' results against the
results published in the paper.

\subsection{Case Study 2: \textbf{Enc+MLP}}
\label{subsec:case_encmlp}

Enc + MLP is proposed by Chen et al.~\cite{Chen:USENIX23} to
address the concept drift~\cite{Jordaney:USENIX17} problem in Android
malware detection. This scheme integrates contrastive learning with
active learning, structuring the model into two subnetworks. The first
subnetwork, a hierarchical contrastive encoder
(\textbf{\textit{Enc}}), employs contrastive learning techniques to
encode input embeddings, ensuring that embeddings of the same malware
family are closer to each other. The second subnetwork is an
\textbf{MLP} classifier that utilizes these embeddings for malware
classification.

The model is initially trained on a labeled dataset to establish the
hierarchical contrastive classifier. Following this, an active
learning process begins, where the trained classifier predicts labels
for a batch of test samples. A pseudo-loss sample selector in the
hierarchical contrastive classifier identifies the most uncertain apps
within a predefined labeling budget and adds them to the training set
for the next iteration. The model is then retrained using a warm
start, and this process repeats continuously, refining the model with
each subsequent batch of test data. This approach improves upon
existing active learning baselines for Android malware detection while
minimizing the need for manual labeling, as demonstrated through
evaluations on two publicly available datasets.

\subsubsection{Benchmarking for Enc+MLP}

Enc+MLP uses multiple baseline models and explores
uncertainty sampling for both binary and multiclass classifiers.  The
binary classifiers include a fully connected neural network (Binary
MLP), a linear Support Vector Machine (Binary SVM), and Gradient
Boosted Decision Trees (GBDT).  The multiclass classifiers include an
MLP and an SVM, and the authors also experimented with a combined
classifier, ``Multiclass MLP + Binary SVM.''  Uncertainty is measured as
one minus the maximum prediction score across all classes.  All
baseline models employ a cold-start approach (retraining from scratch
on the updated training data) for active learning.

\subsubsection{Datasets for Enc+MLP}

Enc+MLP was evaluated on two datasets: APIGraph and AndroZoo.\\

\begin{table}[b]
  \centering
        \begin{tabular}{c|c|c|c}
        \bottomrule
        \textbf{Year} & \multicolumn{1}{c|}{\begin{tabular}[c|]{@{}c@{}}\textbf{Malicious}\\ \textbf{Apps}\end{tabular}} & \multicolumn{1}{c|}{\begin{tabular}[c]{@{}c@{}}\textbf{Benign} \\ \textbf{Apps}\end{tabular}} & \textbf{Total} \\ 
        \hline
        2012 & 3,061  & 27,472  & 30,533  \\ 
        2013 & 4,854  & 43,714  & 48,568  \\ 
        2014 & 5,809  & 52,676  & 58,485  \\ 
        2015 & 5,508  & 51,944  & 57,452  \\ 
        2016 & 5,324  & 50,712  & 56,036  \\ 
        2017 & 2,465  & 24,847  & 27,312  \\ 
        2018 & 3,783  & 38,146  & 41,929  \\ 
        \bottomrule
    \end{tabular}
    \caption{APIGraph Dataset}
    \label{tab:apigraph}
  \end{table}

 \begin{table}[b]
   \centering
        \begin{tabular}{c|c|c|c}
        \bottomrule
        \textbf{Year} & \multicolumn{1}{c|}{\begin{tabular}[c|]{@{}c@{}}\textbf{Malicious}\\ \textbf{Apps}\end{tabular}} & \multicolumn{1}{c|}{\begin{tabular}[c]{@{}c@{}}\textbf{Benign} \\ \textbf{Apps}\end{tabular}} & \textbf{Total} \\ 
        \hline
        2019 & 4,542  & 40,947  & 45,489  \\ 
        2020 & 3,982  & 34,921  & 38,904  \\ 
        2021 & 1,676  & 13,985  & 15,662  \\ 
        \bottomrule
    \end{tabular}
    \caption{AndroZoo Dataset}
    \label{tab:androzoo}   
 \end{table}

\textbf{APIGraph Dataset.} Collected by Chen et
al.~\cite{Chen:USENIX23} using the app hash list provided by
APIGraph~\cite{Zhang:CCS20}, this dataset spans seven years of Android
apps from 2012 to 2018. The apps are ordered based on their appearance
timestamps in VirusTotal~\cite{VirusTotal}, addressing both spatial
and temporal biases~\cite{Pendlebury:USENIXSecurity19}. The dataset is
evenly distributed across the years, with each year consisting of
approximately 90\% benign apps and 10\% malicious apps. The malicious
apps are sourced from VirusTotal~\cite{VirusTotal},
VirusShare~\cite{VirusShare}, and the AMD dataset~\cite{AMD:dimva17},
while the benign apps are obtained from
AndroZoo~\cite{Allix:MSR16}. Table~\ref{tab:apigraph} provides a
detailed breakdown of this dataset.


\begin{table*}[hb]
  \centering
    \begin{tabular}{c|c|c|ccc|ccc}
      \toprule[0.5pt]
      \multirow{3}{*}{\begin{tabular}[c]{@{}c@{}}\textbf{Monthly}\\ \textbf{Sample}\\ \textbf{Budget}\end{tabular}} & \multirow{3}{*}{\begin{tabular}[c]{@{}c@{}}\textbf{Model}\\ \textbf{Architecture}\end{tabular}} & \multirow{3}{*}{\begin{tabular}[c]{@{}c@{}}\textbf{Sample}\\ \textbf{Selector}\end{tabular}} & \multicolumn{3}{c|}{\multirow{2}{*}{\begin{tabular}[c]{@{}c@{}}\textbf{APIGraph Dataset}\\ Average Performance (\%)\end{tabular}}} & \multicolumn{3}{c}{\multirow{2}{*}{\begin{tabular}[c]{@{}c@{}}\textbf{AndroZoo Dataset}\\ Average Performance (\%)\end{tabular}}}\\                   & & & \multicolumn{3}{c|}{} & \multicolumn{3}{c}{} \\
      & & & FNR & FPR & F1 & FNR & FPR & F1 \\
   
        \hline\hline

        \multirow{10}{*}{50}                                                                               & Binary MLP & Uncertainty & 23.77 & 0.52 & 83.84 & 53.12 & 0.46 & 59.50 \\
        \cline{2-9}                                                                                       & Multiclass MLP & Uncertainty & 16.10 & 4.64 & 73.77 & 49.86 & 28.52 & 28.65 \\
        \cline{2-9}                                                                                       & \begin{tabular}[c]{@{}c@{}}Multiclass MLP\\ + Binary SVM\end{tabular} & Uncertainty & 38.40 & 1.01 & 71.38 & 73.13 & 2.87 & 34.04 \\
        \cline{2-9}
        & \multirow{2}{*}{Binary SVM} & Uncertainty & \textbf{16.92} & \textbf{0.61} & \textbf{87.72} & \textbf{48.77} & \textbf{0.29} & \textbf{63.42}  \\
        &  & CADE OOD & 36.11 & 12.9 & 71.70 & 62.01 & 0.55 & 50.26 \\
        \cline{2-9}                                                                                       & multiclass SVM & Uncertainty & 35.79 & 0.17 & 87.43 & 65.77 & 0.09 & 46.91 \\
        \cline{2-9}
        & multiclass GBDT & Uncertainty & 31.75 & 0.54 & 77.92 & 50.35 & 0.47 & 61.06 \\
        \cline{2-9}
        & Enc + MLP & Pseudo Loss & \textbf{15.15} & \textbf{0.64} & \textbf{89.39} & \textbf{27.65} & \textbf{0.53} & \textbf{79.92}  \\
        \cline{2-9}                                                                                       & \cellcolor{gray!20} RF & \cellcolor{gray!20} Uncertainty & \cellcolor{gray!20} 17.99 & \cellcolor{gray!20} 0.20 & \cellcolor{gray!20} 88.95 & \cellcolor{gray!20} 50.28 & \cellcolor{gray!20} 0.10 & \cellcolor{gray!20} 61.87 \\
        \cline{2-9}                                                                                       &  \cellcolor{gray!20} CatBoost & \cellcolor{gray!20} Uncertainty & \cellcolor{gray!20} \textbf{14.70} & \cellcolor{gray!20} \textbf{0.33} & \cellcolor{gray!20} \textbf{90.38} & \cellcolor{gray!20} \textbf{25.72} & \cellcolor{gray!20} \textbf{0.14} & \cellcolor{gray!20} \textbf{83.81} \\
        \hline\hline

        \multirow{10}{*}{100}                                                                              & Binary MLP & Uncertainty & 20.64 & 0.49 & 86.03 & 46.39 & 0.30 & 65.26 \\
        \cline{2-9}                                                                                       & Multiclass MLP & Uncertainty & 14.77 & 6.44 & 69.91 & 35.34 & 32.64 & 33.72 \\
        \cline{2-9}                                                                                       & \begin{tabular}[c]{@{}c@{}}Multiclass MLP\\ + Binary SVM\end{tabular} & Uncertainty & 30.45 & 1.76 & 74.11 & 73.47 & 3.88 & 31.69 \\
        \cline{2-9}
        & \multirow{2}{*}{Binary SVM} & Uncertainty & \textbf{15.41} & \textbf{0.68} & \textbf{88.38} & \textbf{43.07} & \textbf{0.32} & \textbf{68.33}  \\
        &  & CADE OOD & 23.48 & 0.96 & 82.22 & 58.78 & 0.70 & 52.47 \\
        \cline{2-9}                                                                                       & multiclass SVM & Uncertainty & 23.86 & 0.17 & 82.18 & 54.29 & 0.12 & 58.26 \\
        \cline{2-9}
        & multiclass GBDT & Uncertainty & 27.76 & 0.67 & 80.15 & 48.59 & 0.76 & 62.58 \\
        \cline{2-9}
        & Enc + MLP & Pseudo Loss & \textbf{13.69} & \textbf{0.44} & \textbf{90.42} & \textbf{27.35} & \textbf{0.41} & \textbf{80.07}  \\     
        \cline{2-9}                                                                                       & \cellcolor{gray!20} RF & \cellcolor{gray!20} Uncertainty & \cellcolor{gray!20} 15.38 & \cellcolor{gray!20} 0.20 & \cellcolor{gray!20} 90.53 & \cellcolor{gray!20} 48.59 & \cellcolor{gray!20} 0.09 & \cellcolor{gray!20} 63.64 \\
        \cline{2-9}                                                                                       &  \cellcolor{gray!20} CatBoost & \cellcolor{gray!20} Uncertainty & \cellcolor{gray!20} \textbf{11.77} & \cellcolor{gray!20} \textbf{0.26} & \cellcolor{gray!20} \textbf{92.39} & \cellcolor{gray!20} \textbf{26.63} & \cellcolor{gray!20} \textbf{0.13} & \cellcolor{gray!20} \textbf{82.12} \\
        \hline\hline

        \multirow{10}{*}{200}                                                                              & Binary MLP & Uncertainty & 19.71 & 0.39 & 86.97 & 42.57 & 0.34 & 68.47 \\
        \cline{2-9}                                                                                       & Multiclass MLP & Uncertainty & 14.56 & 4.26 & 75.65 & 39.78 & 34.76 & 28.59 \\
        \cline{2-9}                                                                                       & \begin{tabular}[c]{@{}c@{}}Multiclass MLP\\ + Binary SVM\end{tabular} & Uncertainty & 29.46 & 1.98 & 74.09 & 70.32 & 0.93 & 39.51 \\
        \cline{2-9}
        & \multirow{2}{*}{Binary SVM} & Uncertainty & \textbf{14.07} & \textbf{0.86} & \textbf{88.47} & \textbf{40.31} & \textbf{0.37} & \textbf{70.24}  \\
        &  & CADE OOD & 21.68 & 0.67 & 84.50 & 51.32 & 0.78 & 59.11 \\
        \cline{2-9}                                                                                       & multiclass SVM & Uncertainty & 21.19 & 0.21 & 86.90 & 44.77 & 0.13 & 66.55 \\
        \cline{2-9}
        & multiclass GBDT & Uncertainty & 24.71 & 0.56 & 82.71 & 42.97 & 0.80 & 67.28 \\
        \cline{2-9}
        & Enc + MLP & Pseudo Loss & \textbf{9.42} & \textbf{0.48} & \textbf{92.72} & \textbf{27.67} & \textbf{0.39} & \textbf{80.51}  \\
        \cline{2-9}                                                                                       & \cellcolor{gray!20} RF & \cellcolor{gray!20} Uncertainty & \cellcolor{gray!20} 13.44 & \cellcolor{gray!20} 0.19 & \cellcolor{gray!20} 91.76 & \cellcolor{gray!20} 49.45 & \cellcolor{gray!20} 0.90 & \cellcolor{gray!20} 62.74 \\
        \cline{2-9}                                                                                       &  \cellcolor{gray!20} CatBoost & \cellcolor{gray!20} Uncertainty & \cellcolor{gray!20} \textbf{10.40} & \cellcolor{gray!20} \textbf{0.24} & \cellcolor{gray!20} \textbf{93.27} & \cellcolor{gray!20} \textbf{22.19} & \cellcolor{gray!20} \textbf{0.13} & \cellcolor{gray!20} \textbf{86.06} \\
        \hline\hline

        \multirow{10}{*}{400}                                                                              & Binary MLP & Uncertainty & \textbf{16.04} & \textbf{0.40} & \textbf{89.25} & 36.25 & 0.34 & 73.70 \\
        \cline{2-9}                                                                                       & Multiclass MLP & Uncertainty & 15.07 & 4.15 & 75.94 & 34.48 & 24.44 & 38.34 \\
        \cline{2-9}                                                                                       & \begin{tabular}[c]{@{}c@{}}Multiclass MLP\\ + Binary SVM\end{tabular} & Uncertainty & 28.85 & 1.68 & 75.69 & 73.94 & 1.92 & 33.74 \\
        \cline{2-9}
        & \multirow{2}{*}{Binary SVM} & Uncertainty & 12.86 & 0.90 & 89.02 & 34.73 & 0.43 & 74.12  \\
        &  & CADE OOD & 20.61 & 0.59 & 85.52 & 49.98 & 0.94 & 59.53 \\
        \cline{2-9}                                                                                       & multiclass SVM & Uncertainty & 17.87 & 0.24 & 88.88 & 40.99 & 0.14 & 69.61 \\
        \cline{2-9}
        & multiclass GBDT & Uncertainty & 20.16 & 0.46 & 86.24 & \textbf{33.62} & \textbf{0.38} & \textbf{76.82} \\
        \cline{2-9}
        & Enc + MLP & Pseudo Loss & \textbf{7.84} & \textbf{0.50} & \textbf{93.50} & \textbf{21.49} & \textbf{0.31} & \textbf{85.81}  \\
        \cline{2-9}                                                                                       & \cellcolor{gray!20} RF & \cellcolor{gray!20} Uncertainty & \cellcolor{gray!20} 11.06 & \cellcolor{gray!20} 0.20 & \cellcolor{gray!20} 93.13 & \cellcolor{gray!20} 47.79 & \cellcolor{gray!20} 0.90 & \cellcolor{gray!20} 64.29 \\
        \cline{2-9}                                                                                       &  \cellcolor{gray!20} CatBoost & \cellcolor{gray!20} Uncertainty & \cellcolor{gray!20} \textbf{9.41} & \cellcolor{gray!20} \textbf{0.23} & \cellcolor{gray!20} \textbf{93.87} & \cellcolor{gray!20} 26.60 & \cellcolor{gray!20} 0.10 & \cellcolor{gray!20} 82.16 \\
        \hline
    \end{tabular}
    \caption{Performance comparison of different models on APIGraph and AndroZoo datasets.  The unshaded results are from the original paper, while the  shaded results are obtained as part of our benchmarking.}
    \label{tab:chen_dataset_comparison}
\end{table*}

\textbf{AndroZoo Dataset.} Collected by Chen et
al.~\cite{Chen:USENIX23} from AndroZoo~\cite{Allix:MSR16}, this
dataset includes Android apps from 2019 to 2021.The apps are also
ordered based on their appearance timestamps. Malicious apps were
randomly selected based on VirusTotal reports, where at least 15
antivirus engines flagged them as malware. Benign apps were randomly
chosen from those with 0 positive detections by any antivirus
engine. The dataset maintains a similar malware/benign app ratio, with
90\% benign apps and 10\% malicious apps each
year. Table~\ref{tab:androzoo} provides a detailed breakdown of the
collected data for each year.

\subsubsection{Additional Benchmarking}


Enc + MLP was evaluated on the two datasets and compared with multiple
baseline models.  The results showed that it outperforms all baseline
models, with a particularly significant improvement on the AndroZoo
dataset. The authors have made all research artifacts publicly
available, including the datasets used in the experiments and the
model-related code. This accessibility allows us to integrate
additional testing models into their framework seamlessly, enabling
the evaluation of more ML models using the same input and evaluation
metrics.

We selected additional ML models from the ones mentioned in
Section~\ref{subsec:traditional_ml}: NB, KNN, RF, and CatBoost
(excluding SVM, as it was already included as a baseline in the
original work), to expand the set of baseline
models. Table~\ref{tab:chen_dataset_comparison} presents the
performance of all tested models, with shaded rows indicating models
not included in the original work. If an additional model did not
outperform the original baseline models, we omitted its results from
the table. As a result, Table~\ref{tab:chen_dataset_comparison} only
includes RF and CatBoost as additional baseline models. The results
show that RF achieves performance comparable to the Enc+MLP model on
the APIGraph dataset but performs poorly on the AndroZoo
dataset. However, CatBoost outperforms Enc+MLP on both datasets across
different labeling budgets, except for the AndroZoo dataset when the
labeling budget is set to 400.

\subsection{Case Study 3: \textbf{CapsGNN}}
\label{subsec:case_capsgnn}

\textbf{Capsule Graph Neural Networks.}  The Capsule Graph Neural
Network (CapsGNN) approach proposed by
Zhang~et~al.~\cite{Zhang:ICLR19} represents an attractive approach to
explore in the context of Android malware detection, since it captures
information at graph level and enhances graph embeddings obtained with
Graph Neural Networks (GNN)~\cite{Kipf:LCLR17} by adopting the Capsule
Neural Network (CapsNet)~\cite{Sabour:arXiv17} ideas. CapsNet helps
capture different aspects of the graph, corresponding to different
properties, in the final embeddings.

The Capsule Neural Networks (CapsNet) architecture was introduced by
Sabour~et~al.~\cite{Sabour:arXiv17} for image feature extraction,
although the core concept of a capsule in neural networks was created
by Hinton et al.~\cite{Hinton:ICANN2011}. The CapsNet's design is
based on CNNs, but in addition to detecting features as a CNN does,
CapsNet also aims to encode instantiation parameters (such as
position, orientation, texture) as part of the detected
features. Thus, rather than representing the features using scalar
values as in CNNs, the CapsNet uses capsules, which represent a group
of neurons, to encode features as vectors.  The length of a capsule
determines the probability that the corresponding feature is
present. The direction of the capsules reflects the instantitation
parameters of the feature.

Rather than using a pooling layer to transmit the feature information
between layers, CapsNet uses a dynamic routing mechanism to find the
best connections between the current and next layer capsules based on
agreement. The key innovation of routing by agreement is the
capability to capture the spatial relationships between parts and
their whole. In our case, the CapsNet is meant to encode precise
semantic information from the inter-procedural control flow
graphs (ICFGs) of an app.

\begin{figure}[t]
   \centering
   \includegraphics[width=\linewidth]{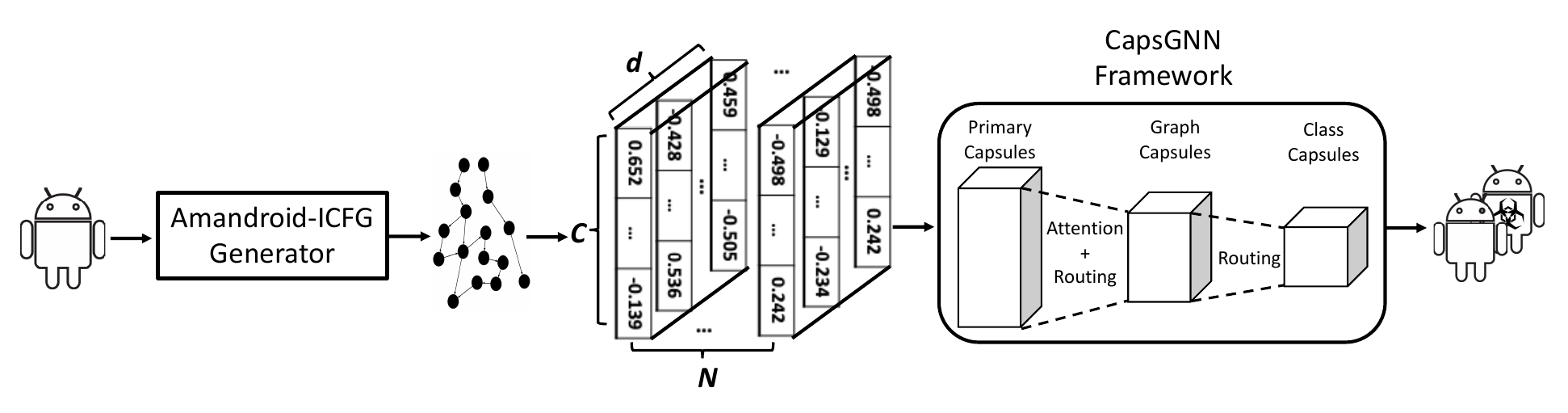}
   \caption{CapsGNN Architecture in Android Malware Detection. \(\textbf{N}\) is the number of nodes in one graph, \(\textbf{C}\) is the number of node attribute channels, \(\textit{d}\) is the node embedding dimension.}
   \label{fig:capsgnn_view}
   \vspace{-.2in}
\end{figure}

As GCN, CapsGNN takes as input a graph data structure, while also
employing capsules (vectors) instead of scalar values to represent
different instantiation parameters of the features. Thus, CapsGNN is a
promising DL candidate for Android malware detection, as it can work
directly with the ICFG and capture different aspects of the
graphs. Figure~\ref{fig:capsgnn_view} illustrates the CappsGNN
architecture implemented in our Android malware detection
system. CapsGNN extracts a primary capsule for each node in an input
graph. As opposed to GCN, which extracts node embeddings from the last
layer of the network, CapsGNN extracts node embeddings from different
layers and represents them as primary capsules (more precisely, the
primary capsules are obtained by summing up node embeddings from
different layers). As graphs in other application domains, the ICFGs
generated from Android apps are diverse, with graph sizes varying
widely. As mentioned above, the number of primary capsules depends on
the number of nodes in a graph, as each capsule corresponds to a
node. Furthermore, the number of graph and class capsules depends on
the number of primary capsules. Thus, the node capsules need to be
scaled to ensure that the graph capsules from different graphs have
the same dimension and are comparable, despite differences in graph
sizes. The CapsGNN framework designed by
Zhang~et~al.~\cite{Zhang:ICLR19}, which we use in this work,
implements an attention module between primary capsules and graph
capsules. The attention module ensures that: 1) the model identifies
the most relevant parts of the graph, and 2) the node capsules are
scaled to the same dimension, so that they subsequently lead to
comparable graph and class embeddings.

\subsubsection{Benchmarking for CapsGNN}

CapsGNN is a DL model designed to capture complex patterns within an
app. We include five widely used ML models, as mentioned in
Section~\ref{subsec:traditional_ml}, as potential baselines for
comparison with the CapsGNN model.

In addition to traditional ML models, we also experimented with
ExcelFormer, a state-of-the-art neural network proposed by Chen et
al.~\cite{excelformer:kdd24}, which outperforms Gradient Boosted
Decision Trees (GBDTs) and existing DL models for tabular data. It
addresses key challenges in deep tabular learning through three key
innovations: (1) a semi-permeable attention module that reduces the
influence of less informative features, (2) data augmentation
techniques specifically designed for tabular data, and (3) an
attentive feedforward network that enhances model fitting
capability. These design choices make ExcelFormer a highly effective
solution for diverse tabular datasets. Extensive and rigorous
experiments on real-world datasets demonstrate that ExcelFormer
consistently outperforms previous models across various tabular
prediction tasks. Given that an APK's structural representations can
be effectively converted into tabular data, we selected ExcelFormer to
assess its potential for improving Android malware detection
performance.

The Android system is an event-based system, and the event-driven
control flow can involve various method calls based on the app's
components' life cycles. Independent method-level control flow graphs
(CFGs) or API call sequences could not capture the true invocation
order of API calls. In contrast, a ICFG provides a more accurate
representation of the actual execution sequence of
these. operations. Therefore, we utilize ICFGs as the input
representation for our CapsGNN model.

Traditional ML models usually need a predefined feature set to
represent an application, and the performance of the classifier highly
depends on the features. Section~\ref{sec:related} briefly explains
some prior works' approaches to feature selection. The most common
features used in prior works are API calls, app permissions, and a few
other pieces of information from raw app code. The feature sources are
similar but different approaches collect and organize the features in
different ways. Among prior works based on traidtional ML,
DREBIN~\cite{Arp:NDSS14} has shown high detection preformance on
datasets with realistic size and malware to benign app ratio. Recent
work~\cite{Daoudi:TOPS22} shows that DREBIN's features are still
robust for apps as late as 2019. DREBIN extracts static features such
as components, intents, API calls, permissions, and network addresses
totaling 500,000 features. Roy et al.~\cite{Roy:ACSAC15} achieved
similar performance as DREBIN by choosing only 471 static features
from the original DREBIN's feature set. To balance the computational
cost and performance on large-scale datasets, we use the same features
as those in Roy~et~al.'s work~\cite{Roy:ACSAC15}, which consist of 471
features including permissions, intent actions, discriminative APIs,
obfuscation signatures, and native code signatures. To support future
research, the complete feature list will release as open-source with
other related code following the acceptance of this work.

\subsubsection{Dataset for CapsGNN}

\begin{figure*}[t]
     \centering
     \begin{subfigure}[b]{0.38\textwidth}
         \centering
         \includegraphics[width=\textwidth]{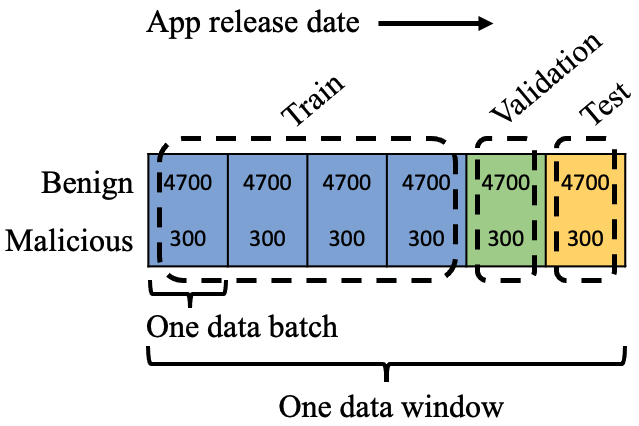}
         \caption{Breakdown of a 6-batch window}
         \label{fig:window_detail}
     \end{subfigure}
     \hfill
     \begin{subfigure}[b]{0.49\textwidth}
         \centering
         \includegraphics[width=\textwidth]{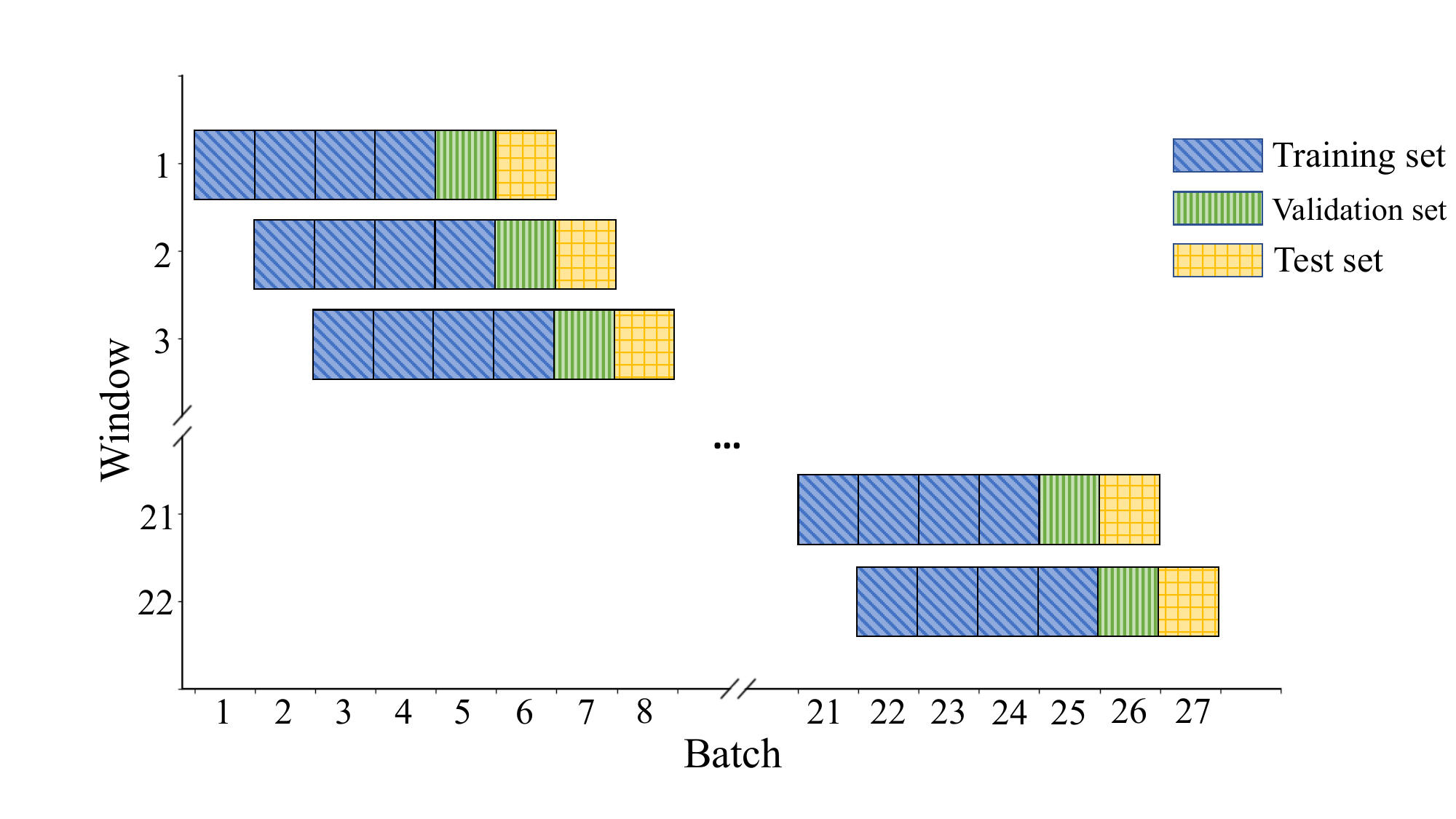}
         \caption{6-batch sliding windows  (with 4 training batches)}
         \label{fig:sliding_window_overview}
     \end{subfigure}
    \hfill
     \begin{subfigure}[b]{0.49\textwidth}
         \centering
         \includegraphics[width=\textwidth]{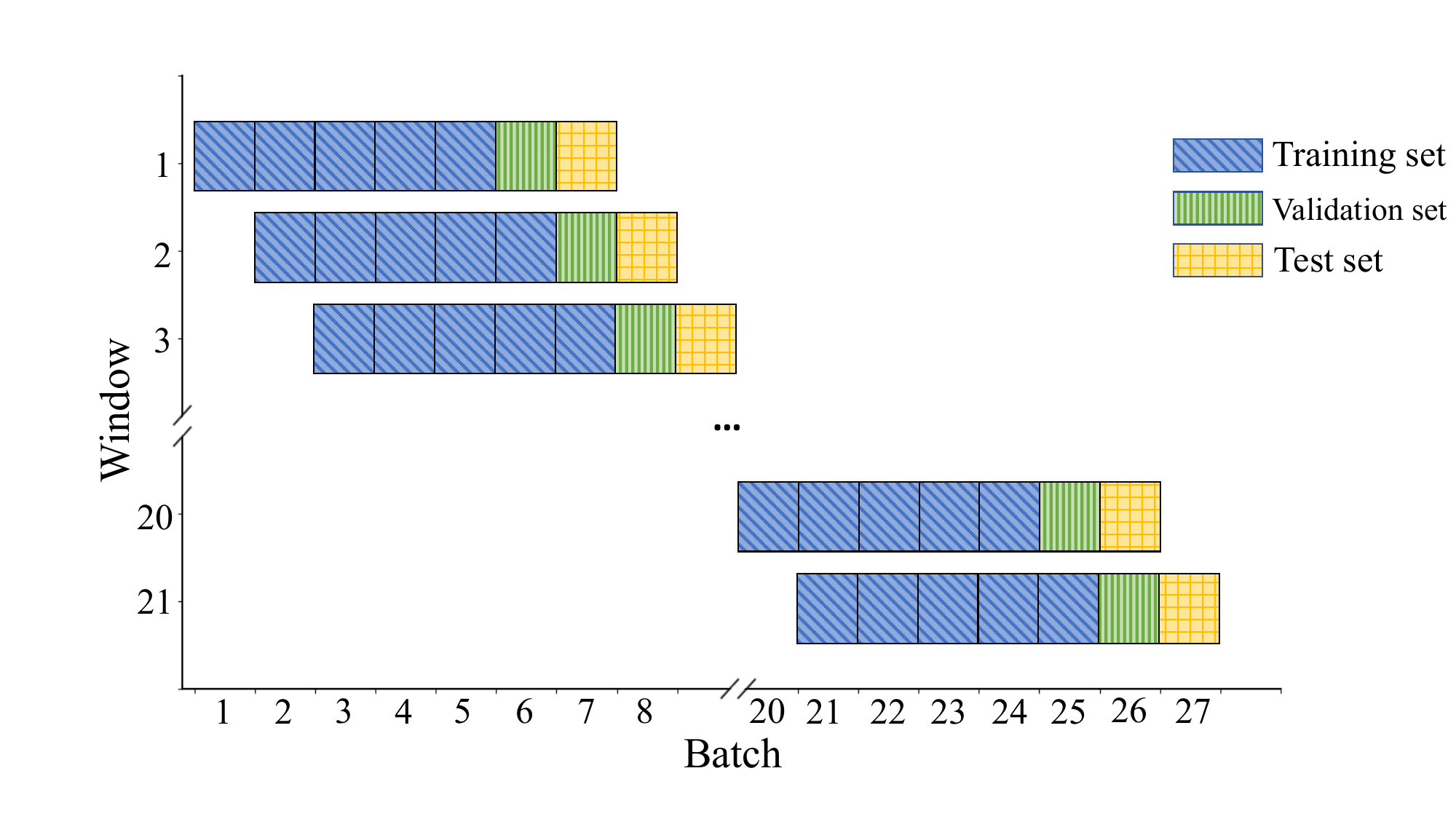}
         \caption{7-batch sliding windows  (with 5 training batches)}
         \label{fig:5_batches}
     \end{subfigure}
     \hfill
     \begin{subfigure}[b]{0.49\textwidth}
         \centering
         \includegraphics[width=\textwidth]{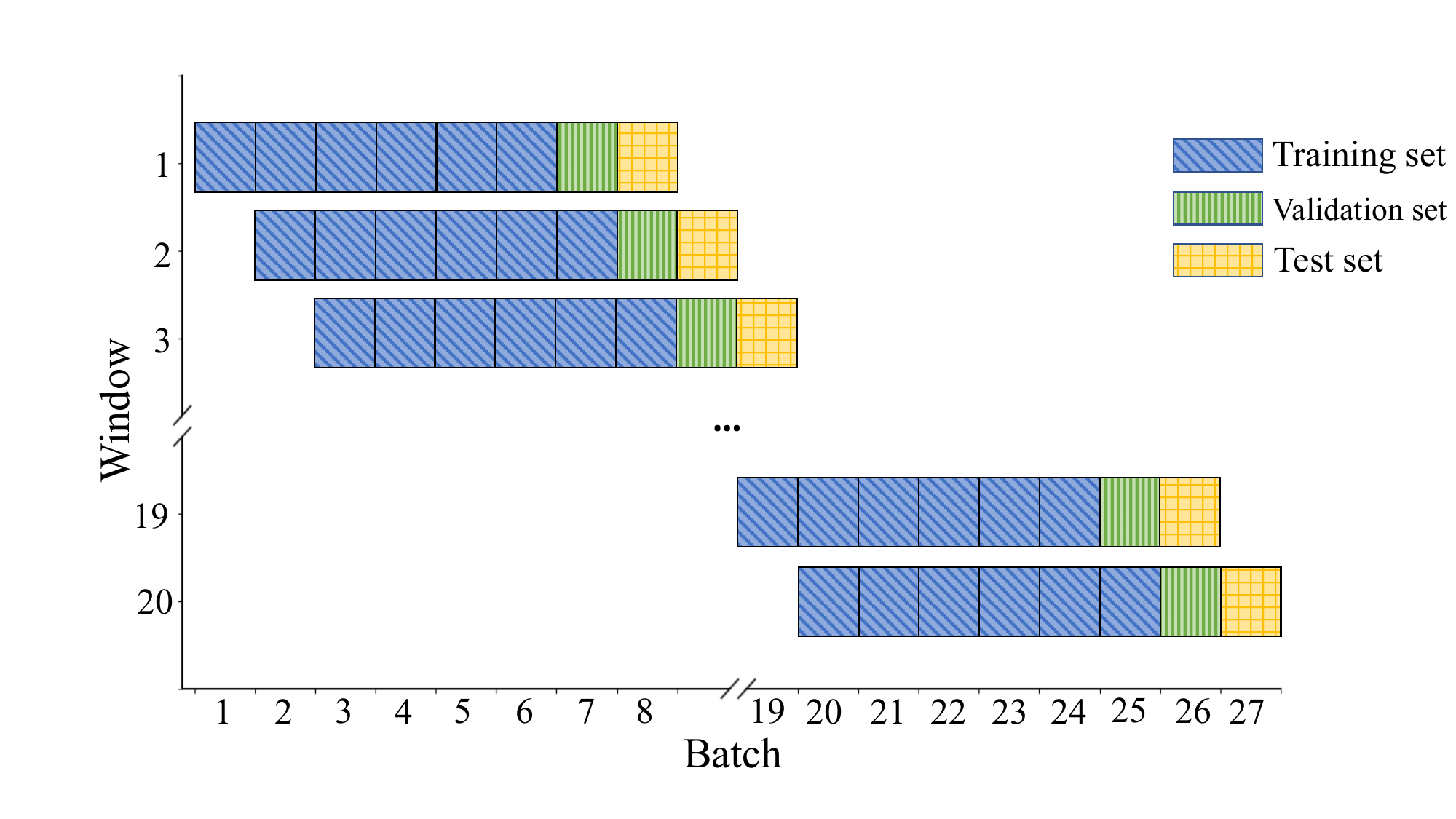}
         \caption{8-batch sliding windows  (with 6 training batches)}
         \label{fig:6_batches}
     \end{subfigure}
        \caption{Sliding Windows}
        \label{fig:Sliding_windows}
 \end{figure*}

The dataset plays a crucial role in ML-based malware detection
research. To ensure meaningful evaluation, the dataset size needs to
reflect the scale of Android apps found in real-world markets. A key
challenge in existing datasets is sampling
bias~\cite{Daniel:USENIXSecurity22}, which can arise when benign and
malicious apps are collected from different sources. To mitigate this
issue, data should be gathered from a single market, ensuring
consistency in distribution and characteristics.  Furthermore, the
organization of training and testing subsets should align with
real-world deployment scenarios. Specifically, training should be
performed on earlier data, while testing should be conducted on later
data—a methodology supported by prior
research~\cite{Pendlebury:USENIXSecurity19}. Additionally, the testing
subset must accurately represent the real-world ratio of benign to
malicious apps. Some industry
experts~\cite{Pendlebury:USENIXSecurity19} suggest that the proportion
of malware in app markets is relatively low, approximately 6\%.  Since
class imbalance can significantly impact classifier performance,
particularly in terms of precision and recall, it is crucial to
construct a dataset that mirrors these real-world distributions.  By
systematically addressing these concerns, we aim to create a more
representative and reliable dataset for Android malware detection
research.

To assess the impact of app evolution on model performance, we adopt a
\emph{sliding window} approach for constructing training, validation,
and test datasets, as illustrated in
Figure~\ref{fig:Sliding_windows}. Specifically, we first partition the
data into batches based on app release dates, with each batch
containing 5,000 apps—comprising 300 malicious apps and 4,700 benign
apps within the same time frame.

\begin{figure}[tbp]
  \centering
  \includegraphics[width=.95\linewidth]{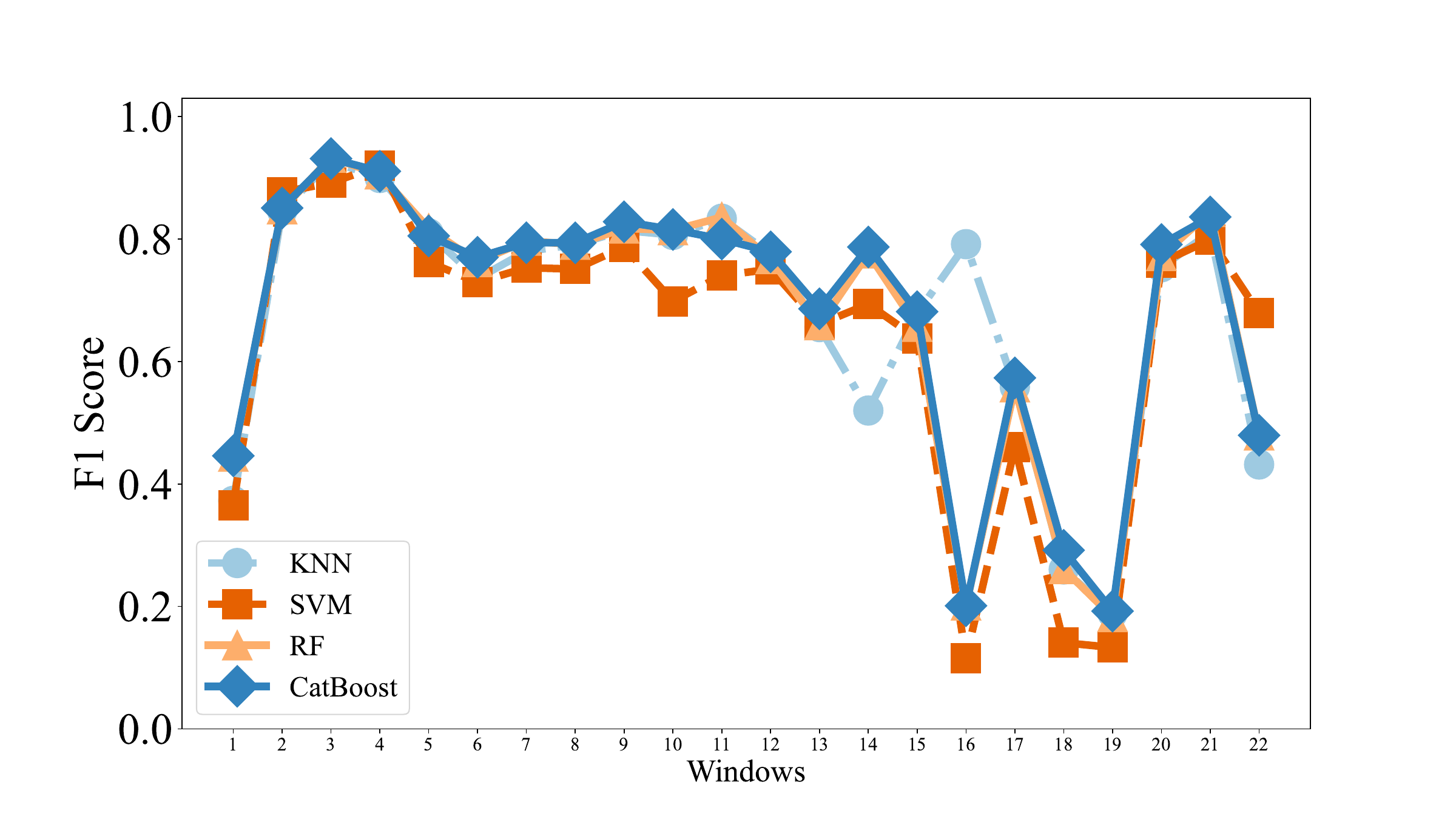}
  \caption{Top-performing models among all tested models}
  \label{fig:top_ml_models}
  \vspace{-.2in}
\end{figure}

\begin{figure*}[b]
    \centering
    \begin{subfigure}{0.47\textwidth}
        \centering
        \includegraphics[width=\textwidth]{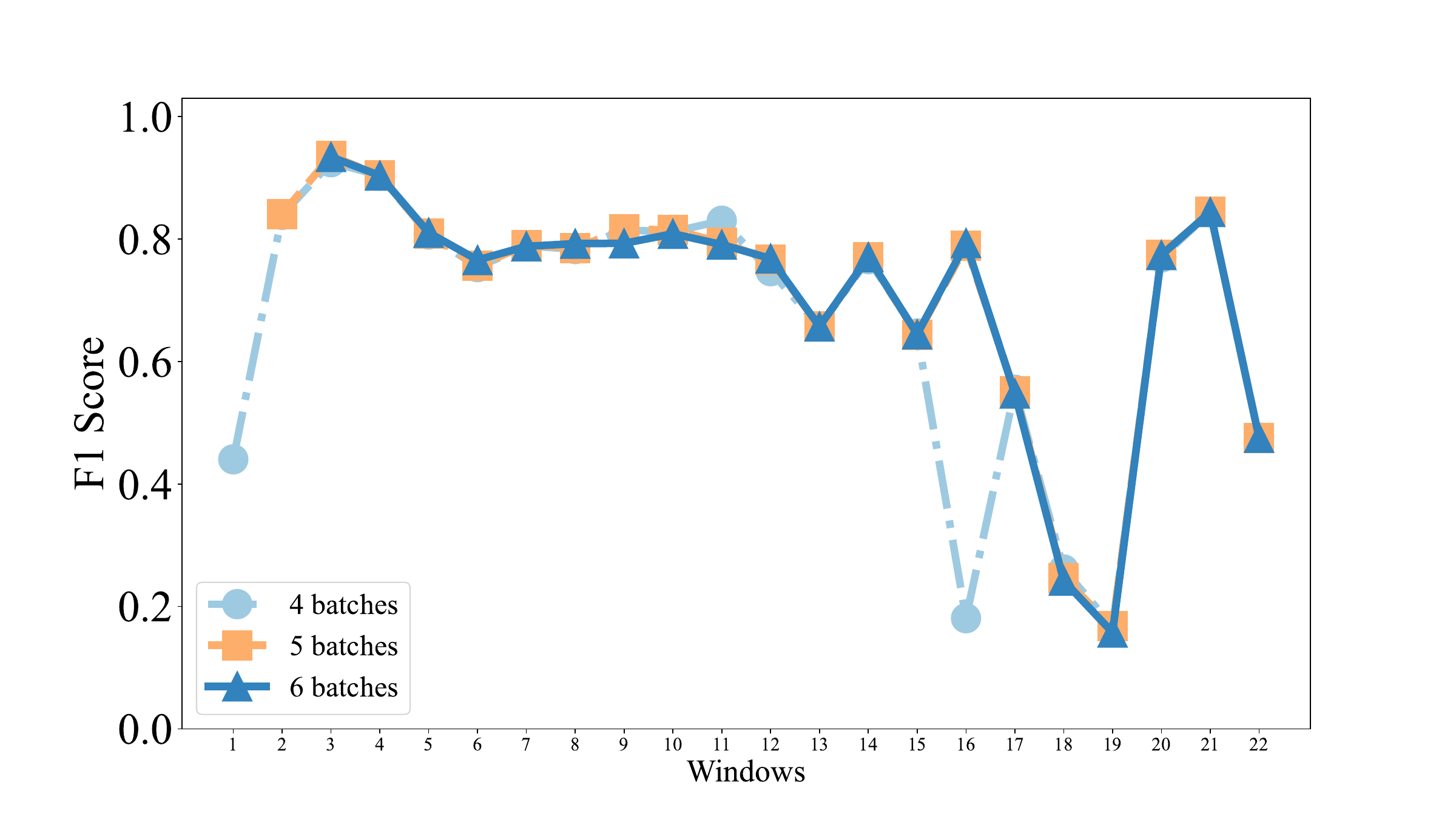}
        \caption{Random forest performance over the sliding windows. The training datasize in a window is varied: 4-batches, 5-batches, and 6-batches.}
        \label{fig:diff_train_rf}
    \end{subfigure}
    \hspace{3mm}
    \begin{subfigure}{0.47\textwidth}
        \centering
        \includegraphics[width=\textwidth]{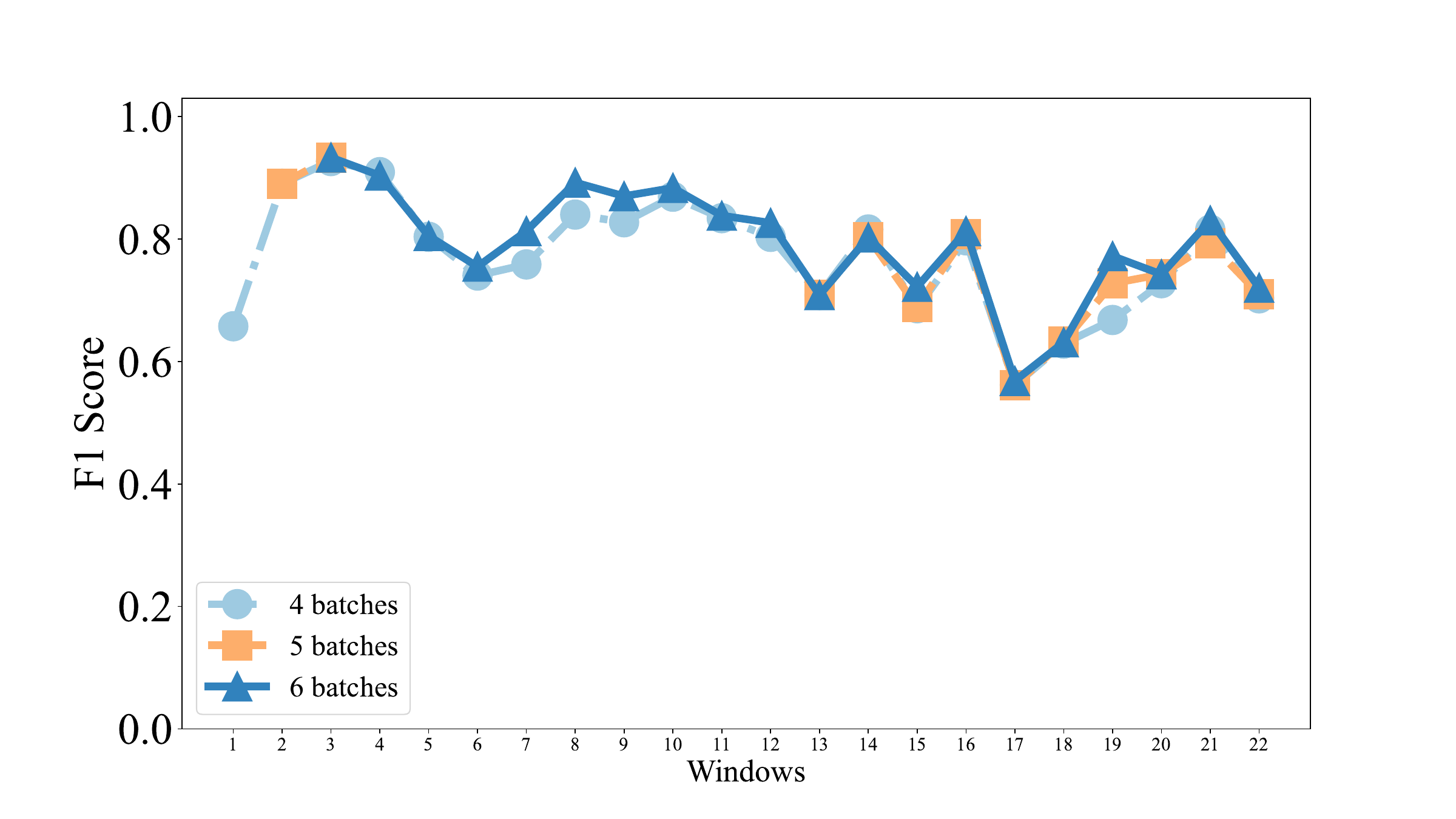}
        \caption{CapsGNN performance over the sliding windows. The training datasize in a window is varied: 4-batches, 5-batches, and 6-batches.}
        \label{fig:diff_train_capsgnn}
    \end{subfigure}
    \caption{RF and CapsGNN performance over the sliding windows}
    \label{fig:main_figure}
\end{figure*}

We subsequently form windows of consecutive batches, with the first
few batches used for training, the batch before the last for
validation and the last batch for testing, as shown in
Figure~\ref{fig:window_detail} for a 6-batch window. In each
subsequent experiment, we slide the window forward by one batch to
form a new dataset for training and evaluation.  Our objective is to
maximize data availability for both training and testing. A larger
training set contributes to more robust models, while an extensive
test set ensures reliable performance estimation. Each batch is
designed to include 5,000 apps, which we believe is sufficient to
provide accurate performance evaluations. To determine the optimal
number of training batches for model stability, we experiment with
windows containing 4, 5, and 6 training batches, as shown in
Figure~\ref{fig:sliding_window_overview}, Figure~\ref{fig:5_batches}
and Figure~\ref{fig:6_batches}, respectively.

As illustrated in Figure\ref{fig:sliding_window_overview}, our dataset
configuration results in 22 overlapping 6-batch windows, with each
window consisting of 28,200 benign apps and 1,800 malicious
apps—except for the final window, where the availability of malicious
apps from late 2021 was limited. Our data partitioning strategy
ensures that all experiments are conducted on subsets of consistent
size and with a fixed malicious-to-benign ratio. Additionally, this
approach enables the observation of temporal trends in Android malware
detection.


We will publicly release this dataset upon the publication of the paper.

\subsubsection{Benchmarking Results}

Figure~\ref{fig:top_ml_models} illustrates the top-performing ML models when using a window with four training batches. It can be observed that all models exhibit very similar performance. RF, CatBoost, and KNN show nearly identical performance, often overlapping on the graph, while SVM performs slightly lower than the other three models across most windows. 


\begin{figure*}[t]
    \centering
    \begin{minipage}{0.5\textwidth}
      \centering
        \includegraphics[width=.9\linewidth]{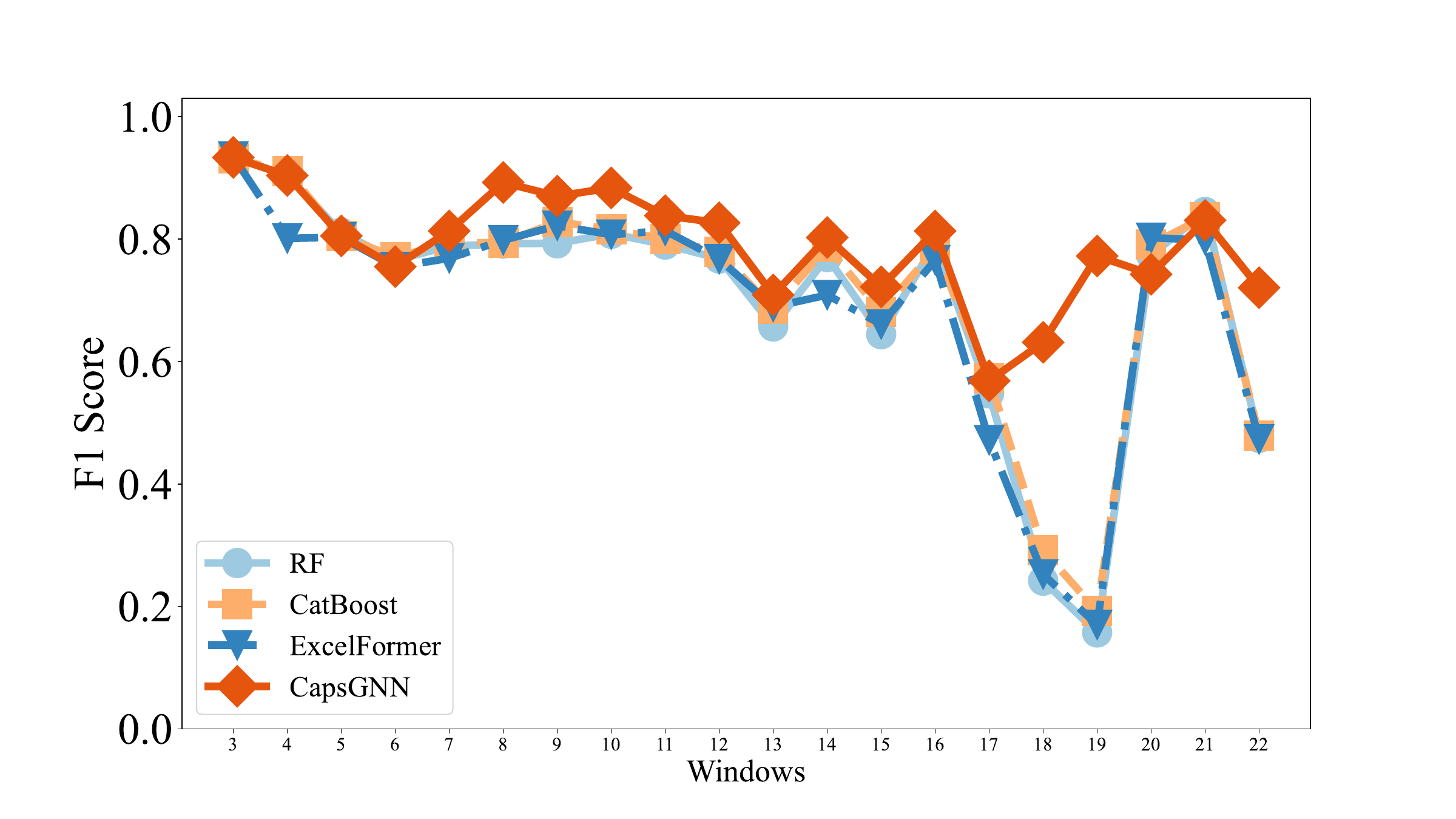}
        \caption{RF, CatBoost, ExcelFormer and CapsGNN Performance Comparison: Training data consists of 6-batches per window.}
        \label{fig:evaluation_result}
    \end{minipage}%
    \hfill
    \begin{minipage}{0.45\textwidth}
      \centering
        \begin{tabular}{l*{2}c}
         \toprule[1pt]
         $_{Size}$ \space \textbackslash \space $^{Model}$
         & \begin{tabular}{@{}c@{}}RandomForest \\ (seconds) \end{tabular}
         &  \begin{tabular}{@{}c@{}}CapsGNN \\ (hours) \end{tabular} \\
         \midrule
         4 batches & 55 & 13 \\
         5 batches & 75 & 17 \\
         6 batches & 89 & 20 \\
         \bottomrule[1pt]
         \end{tabular}
         \vspace{10pt}
        \captionof{table}{Model Running Time with Different Training Data Size}
        \label{tab:model_run_time}
      \end{minipage}
      \vspace{-.2in}
\end{figure*}

\textbf{How much data is needed for model training?} 
To understand how many training batches need to be included in an
experiment window to obtain robust models, we explored windows with
different number of training batches as illustrated in
Figure~\ref{fig:sliding_window_overview}, Figure~\ref{fig:5_batches}
and Figure~\ref{fig:6_batches} , which show windows with 4, 5 and 6
training batches, respectively.

As can be seen in Figure~\ref{fig:diff_train_rf} and
Figure~\ref{fig:diff_train_capsgnn} for RF and CapsGNN, a larger
training dataset does not always result in increased model
performance.

For the traditional ML model, once the size of the training data
reaches a certain level, the performance of the models does not
improve significantly even when we further increase the size of the
training data. In particular, the models trained on 4 batches of data
have similar performance as the models trained on 6 batches of
data. This can be seen for most windows in
Figure~\ref{fig:diff_train_rf}.  For CapsGNN, as
Figure~\ref{fig:diff_train_capsgnn} shows, the performance of the
model improves when training data size in a window is increased from 4
batches to 6 batches (this is expected given that the CapsGNN model
has a large number of parameters that need to be learned). However,
the time and resources required to train the model also prohibitively
increase (see Table~\ref{tab:model_run_time} for time requirements of
the two models), which made it impossible for us to experiment with
even larger training data for CapsGNN.  In contrast, the RF model
needs less than two minutes to train regardless of the size of the
training subset. Given these results, the remaining experiments are
performed using a window size of 8 batches (with 6 batches being used
for training).

The results in Figure~\ref{fig:evaluation_result} show that the
CapsGNN model has similar performance to the other models for windows
from the start of 2018 until the middle of 2019 (i.e., from window 3
through window 14). However, CapsGNN does outperform baseline models
in most test batches past late 2019 (in particular, windows 15, 16,
17, and 18).  We observe that CapsGNN model has more robust
performance than baseline models especially when data evolves (or
undergoes significant changes) past late 2019.  This is not a
surprising result given that the CapsGNN's input includes much richer
program semantics information in the form of ICFG graphs, together
with features not present in the ICFG graph itself, including
permissions, intent actions and obfuscated/native code
signatures. Moreover, CapsGNN dynamically identifies predictive
features in the input as part of the learning process, making it
easier to keep up with changes in the data. As opposed to that, the
input of the RF model includes only manually designed static features
that do not change with changes in the data.  It is surprising to see
that, despite the strictly richer CapsGNN's input and also the
significantly more computational resources required by CapsGNN,
CapsGNN fails to produce significant performance improvements over the
RF model when the data is relatively stable.

\section{Conclusion}
In this paper, we systematically evaluate Android malware detection
models across four datasets—three publicly available and one
systematically collected Google Play-only dataset. We assess the
performance of both deep learning (DL) and traditional machine
learning (ML) models. After implementing additional models for each
proposed DL approach on each dataset, our results show that while DL
models achieve strong performance, simpler and computationally
efficient ML models can achieve comparable or even superior
performance. These findings emphasize the importance of extensive
benchmarking in Android malware detection research. We encourage
future studies to conduct more comprehensive evaluations between
advanced DL models and traditional ML models to provide a more
accurate assessment of detection capabilities. Additionally, we
advocate for releasing datasets used in research to facilitate further
research, enabling deeper exploration and new discoveries in this
field.  We will publicly release our dataset upon the publication of
the paper.

\bibliographystyle{IEEEtran}
\bibliography{android}

\appendix

\subsection{Node Embeddings} \label{app:a}

Our key requirement is to
    feed structural features into the neural networks directly. CapsGNN
    uses a GCN~\cite{Kipf:LCLR17} architecture to obtain node
    embeddings. The GCN is a widely used  architecture,
    which has the capability to work directly on graph data and captures its
    structural information. It is adapted from standard convolutional neural networks (CNN) to graphs.

     A graph with $N$ nodes is represented as a data structure
     containing nodes $V=\{v_1,\cdots,v_N\}$, node
     features $\textbf{X}\in R^{N\times d}$ (where $d$ is the dimension for the node embeddings), and edges
     between nodes encoded as an adjacency matrix $\textbf{A}\in \{0, 1\}^{N \times N}.$ This data structure is provided as input to a GCN with  $L$ convolution layers. At each layer $l$, the convolution operation is applied to each node and its neighbors, or more precisely to the neighbors that can be reached within $l$ steps. Then, the representation of each node at  layer $l$ is obtained by transforming the result of the convolution operation at that layer through an activation function. The  representation at the last layer $L$ is used to obtain the final  embeddings of the nodes in the graph. The dimensionality of the node embeddings is a hyper-parameter that needs to be fine-tuned.
    Formally, the node embeddings at layer $l+1$ are obtained as follows:

       \begin{equation}
        H^{l+1} = f\left({\textbf{D}}^{-\frac{1}{2}}{\textbf{A}}{\textbf{D}}^{-\frac{1}{2}}\textbf{H}^{l}\textbf{W}^{l}\right)
    \end{equation}
    \noindent
    where \(f\left(\cdot\right)\) denotes the activation function; \(H^{l}\) represents the node features at layer $l, H^{0} = \textbf{X}$; 
    \(\textbf{W}^{l}\) is a trainable weight matrix at layer \(l\); \({\textbf{A}}=\textbf{A} + \textbf{I}\), where \(\textbf{I}\) is the identity matrix; 
    \(\textbf{D}\) is the diagonal node degree matrix of \({\textbf{A}}\). The normalized Laplacian matrix defined as \({\textbf{D}}^{-\frac{1}{2}}{\textbf{A}}{\textbf{D}}^{-\frac{1}{2}}\)  helps avoid exploding/vanishing gradients in this deep neural network model.

\subsection{Experimental Setup} \label{app:b}

\textit{Experimental Environment.} Traditional ML models are trained and tested on a Linux server equipped with 48 Intel Xeon(R) E5-2650 cores and 256GB of RAM. For each ML model, we apply a grid search to identify the optimal hyperparameters from a set of predefined candidates. Detailed hyperparameter configurations for each model on different datasets will be provided in the following section. The DL models are trained and tested on a Linux GPU server with 28 Intel(R) Xeon(R) Gold-6132 cores, 512GB of RAM, and a Tesla T4 GPU.

\textit{Evaluation Metrics.}  After running the trained model on a given test dataset, the prediction results can be statistically described in a \textit{confusion matrix} as shown in Table
\ref{tab:confusion_matrix}. To evaluate the models on the GooglePlay-Only dataset, we use three key metrics derived from the confusion matrix: precision, recall, and F1 score. The precision, recall, and F1 score reported in this dataset specifically refer to the \textbf{performance on malicious apps}. Precision (PR) is calculated as \(TP/(TP + FP)\), recall is calculated as \(TP/(TP + FN)\), and F1 score is calculated as \(2 * (Precision * Recall)/(Precision + Recall)\). For publicly available datasets, we follow the same evaluation metrics as used in the original works.

\begin{table}[h]
\caption{Confusion Matrix}
\begin{center}
\begin{tabular}{|c|c|c|c|}
  \hline
  & \multicolumn{3}{c|}{Predict class}\\
  \hline
  \multirow{5}{*}{\begin{tabular}{@{}c@{}}Actual \\ class \end{tabular}} &  & Malicious & Benign\\
  \cline{2-4}
  & \multirow{2}{*}{Malicious} & True Positive & False Negative  \\
  & & (TP) & (FN) \\
  \cline{2-4}
  & \multirow{2}{*}{Bengin} & False Positive & True Negative \\
  & & (FP) & (TN) \\
  \hline
\end{tabular}
\end{center}
\label{tab:confusion_matrix}
\end{table}

\end{document}